\documentclass[preprint,preprintnumbers,amsmath,amssymb]{revtex4-1}

\usepackage{graphicx}
\usepackage{dcolumn}
\usepackage{bm}

\begin{document}
\title{The effect of grain boundaries on the domain wall dynamics in Pr$_{1-x}$Ag$_x$MnO$_3$ manganites}
\author{Hossein Ahmadvand}
\altaffiliation{Corresponding author} \email{ahmadvand@ph.iut.ac.ir}
\author{Hadi Salamati}
\author{Parviz Kameli}
\affiliation{Department of Physics, Isfahan University of Technology, Isfahan 84156-83111, Iran}

\begin{abstract}

{We performed detailed ac susceptibility measurements on Pr$_{1-x}$Ag$_x$MnO$_3$ ($x$=0.15, 0.20) manganites and observed unusual and interesting features, which are associated with the domain walls and the effects of grain-boundaries on their movements. It is shown that the ac field, frequency, temperature, thermal cycling and grain coupling significantly influence the real and especially imaginary parts of the ac susceptibility. We argue that the cooperative depinning of the domain walls from the grain-boundaries accompanying with a large distance movements of the walls leads to the appearance of an anomaly at low temperatures. The anomaly is observed above a threshold ac field and depends on the intergrain coupling. The results show that in the powdered form of bulk samples, the anomaly disappears and the ac field dependence of the $\chi'$ and $\chi''$ suppresses. Below the anomaly temperature, significant and unusual thermal hysteresis occurs in the real and especially imaginary parts of the ac susceptibility. Thermal hysteresis depends on the thermal cycling and indicates thermally irreversible wall pinning and depinning. The susceptibility is practically frequency-independent in the temperature range of thermal hysteresis i.e. below the anomaly. The frequency dependence of ac susceptibility between the anomaly temperature and the transition temperature is discussed by considering the temperature dependence of the relaxation process of the domain walls.}
\end{abstract}

\maketitle
\section{Introduction}

Perovskite manganites (Ln$_{1-x}$A$_{x}$MnO$_3$ ; Ln=La$^{3+}$, Pr$^{3+}$,... ; A=Ca$^{2+}$, Sr$^{2+}$, Ag$^{+}$,...) are a subject of very active research because of their interesting and complex physical properties. Similar to other magnetic materials, magnetic domain walls (DWs) have a strong influence on the magnetic properties of mixed-valence manganites. The DWs also play a significant role in the low-field magnetoresistance of manganites \cite{1,2,3,4,5}. Numerous studies have been devoted to domain/DW effects in manganites, using the direct observation by microscopy methods \cite{2,4,6,7,8,mor,two,sci,nature,l1} and ac susceptibility/magnetization measurements \cite{9,jm,11,12,13,132}. In general, at low applied fields, DW effects are more observable in low doped manganites \cite{9,jm,11,12}. In high doped manganites, such as La$_{0.65}$Ca$_{0.35}$MnO$_3$, the domains are very stable at low temperatures and cannot be easily disturbed \cite{sci}. On the other hand, low doped manganites have a complicated domain structure \cite{7,mor,two}. However, despite the numerous studies, little is known about the DW dynamics in manganites. Especially, accurate investigations concerning the grain - boundaries (GBs) effects on DW dynamics are lacking.

Ac magnetic susceptibility ($\chi=\chi'+i\chi''$) measurements have been widely used for characterizing the various magnetic behaviors such as spin-glass, superparamagnetic and DW dynamics which are occurring in different type of magnetic materials including mixed-valence manganites. The physical mechanisms governing the magnetic response cannot always be distinctly sorted out and thus the analysis of an ac magnetic response can be difficult. For example, Raj Sankar et al. \cite{9} discussed the low temperature anomaly in the ac susceptibility of La$_{1-x}$MnO$_3$
and attributed it to the DWs pinning rather than spin glass. This is because, the frequency dependence of magnetic susceptibility of ferromagnets with a nonzero coercivity and remanence may appear to be similar, in general, to that observed in spin glasses \cite{14}. However, the ac susceptibility is a powerful tool to study DW movements and related effects \cite{15,16,20}. In mixed-valence managanites, appearance of discontinuous jumps \cite{11,13}, disaccommodation phenomenon \cite{12} and low temperature anomaly \cite{9,jm} in the $\chi'$ were explained in terms of DW processes.

Substitution of different ions at Ln/A sites in the perovskite manganites (Ln$_{1-x}$A$_{x}$MnO$_3$) provides a systematic way to study the physical properties and discover interesting effects. Recently, we reported the structural, magnetic and electrical properties of Pr$_{1-x}$Ag$_x$MnO$_3$ ($0\leq x\leq0.25$) manganites \cite{17}. It was reported that the Ag$^{+}$ ion does not substitute for the Pr$^{3+}$ site and the nominal Pr$_{1-x}$Ag$_x$MnO$_3$ can be conveniently described by [Ag + self doped Pr$_{1-y}$MnO$_{3+\delta}$]. For the higher doping contents, the system was found to be ferromagnetic-insulator. In this paper we present a detailed ac susceptibility study of the high doped Pr$_{1-x}$Ag$_x$MnO$_3$ ($x$=0.15, 0.20). We observed significant and unconventional DW effects below the ferromagnetic transition temperature, $T_C$. The Results are interpreted in terms of the dynamics of the DWs in the presence of GBs as the dominant pinning centers. We will show that there is an anomaly at low temperatures which depends on the intergrain coupling.

\section{Experimental}

Polycrystalline samples of Pr$_{1-x}$Ag$_x$MnO$_3$ ($x$=0.15, 0.20) were prepared by the standard solid state reaction route. The details of preparation and characterization of the samples can be found in Ref. \cite{17}. The microstructure studies were performed by scanning electron microscopy (SEM). Temperature dependence of the real ($\chi'$) and imaginary ($\chi''$) parts of ac susceptibility was measured using a Lakeshore 7000 susceptometer. Measurements of susceptibility at different ac fields/frequencies were made in the same run. We measured the ac susceptibility of the samples in both bulk and powder forms. The bulk samples were slab-shaped with typical dimensions of about $1.3\times3\times10$ mm$^{3}$ and a mass density of about 5.7 gr/cm$^{3}$. No demagnetizing corrections were performed. Using the data of Ref. \cite{nd}, the demagnetizing factor was estimated to be about 0.07 (SI) for the bulk samples.

\section{Results and discussion}
\subsection{Field and frequency dependence}

The temperature dependence of the $\chi'$ and $\chi''$ for the bulk sample of Pr$_{0.85}$Ag$_{0.15}$MnO$_3$, measured at different ac fields (1, 3, 5, 7, 10 Oe) with a frequency of 111 Hz, is shown in Fig. 1. Using the $d\chi'/dT$, the $T_{C}$ was found to be 129 K. This is close to the temperature (131.5 K) at which the $\chi''$ deviates from zero and begins to rise. The $\chi'(T)$ and $\chi''(T)$ data exhibit a peak near $T_C$, below which the data shows significant ac field dependence. Especially, the $\chi''(T)$ is more sensitive to the change of ac field amplitude (Fig. 1(b)) and has a relatively large magnitude (Inset Fig. 1). As can be seen in Fig. 1, when the applied ac field increases to about 10 Oe, a bump-like anomaly appears at about $T_a$=105 K in both the $\chi'$ and $\chi''$ curves. This anomaly is also observed at higher silver-doping levels, i.e. for $x$=0.20 and 0.25.

In order to further study the magnetic properties of the samples, the measured bulk sample was well ground manually, then the measurements were repeated on the powder-form sample at the same conditions. The obtained measurements are given in Fig. 1 (labeled as Powder). In the powder measurements, the ac field dependence is much smaller than the bulk sample and the anomaly at $T_a$ also disappears. Moreover, the magnitude of $\chi'$ and $\chi''$ is significantly suppressed, especially in the peaks region, between 100 - 130 K.

Figure 2 shows the frequency dependence of the $\chi'(T)$ and $\chi''(T)$ in an ac field of 10 Oe, for the bulk sample. The anomaly is clearly visible in the $\chi''(T)$ curves. With increasing frequency the magnitude of $\chi'$ and $\chi''$ decreases, which is more pronounced above the anomaly temperature.

In order to interpret the observed magnetic behaviors of Pr$_{1-x}$Ag$_x$MnO$_3$ manganites, we propose a mechanism based on the dynamics of the DWs. In Ref. \cite{17}, we discussed that the Ag$^{+}$ ion does not substitute for the Pr$^{3+}$ site and locate at the GBs in the metallic form. The enhancement of $T_C$ of Ag doped samples was attributed to the oxygen release from the metallic silver at high temperatures. As a result of the oxygenation of manganite grains, ferromagnetic domains may form at the surface of grains. The ferromagnetic domains are assumed to form in a manner that their walls nucleate and pin at the GBs. Metallic silver can also act as pinning centers for the DWs. In fact, the GBs provide potential wells which can trap the DWs. This is because, when a DW intersects the GBs the associated magnetostatic energy of the GBs can be eliminated \cite{pinn}. As a result, the DWs tend to stay pinned at the GBs, and energy is required to move them past the pinning centers. Principally, appearance of nonzero $\chi''$ below $T_C$ of ferromagnets reflects the energy losses due to the DW movements and domain magnetization rotation induced by ac field. The excitation by an ac field produces two effects; oscillation of the DWs within the potential wells with no imaginary part and hopping from well to well which gives rise to both imaginary and real parts \cite{20}. The type of behavior depends on amplitude and frequency of the ac field. On the basis of the above explanations, we can discuss our results. In an ac field of 1 Oe (Fig. 1), the main response of the system originates from the DW hoping between the wells located at the GBs. By increasing temperature, thermal activation of the DWs results in the increasing of the $\chi'$ and $\chi''$ (Fig. 1). By increasing the ac field amplitude, more DWs can overcome their potential wells and consequently above a threshold field, the walls cooperatively depin from the GBs and move for a rather large distance. As a consequence, we attribute the anomaly at $T_a$ to the cooperative movement of large numbers of the depinned DWs from the GBs. This process significantly raises the $\chi'$ and $\chi''$ at around $T_a$. Relatively large magnitude of the $\chi''$ (inset Fig. 1) gives further evidence for activation and contribution of large numbers of the DWs to the ac losses.

The SEM images of the powder and bulk samples are shown in Fig. 3. The average grain size of the bulk sample is below 5 $\mu m$ (Fig. 3(a)). As Fig. 3(b) shows, the manganite grains are almost detached from each other in the powder sample. Grinding the bulk sample into free grains can change the magnetic characteristics of the sample. Nevertheless, it can be understood from Fig. 1 that the anomaly is not intrinsic. We will show that there is a correlation between the anomaly and intergrain coupling, which is evidence that the GBs are responsible for the appearance of the anomaly. It should be mentioned that the powdering of a bulk ferromagnetic material can lead to the change of the effective shape anisotropy (demagnetization factor) and can thus affect the magnitude of susceptibility.

The frequency dependence of the $\chi'$ and $\chi''$ can be interpreted in terms of relaxation processes of the DWs. In the simplest case only one excitation energy is necessary to explain the thermal activation of the DWs by the applied ac field. Some authors have used the Arrhenius law ($\tau=\tau_0exp(E_a/kT)$) for the temperature dependence of the relaxation time \cite{14,15,16}. Where $\tau$ and $E_a$ are the relaxation time and activation energy, respectively. According to the approach of Chen et al. \cite{15}, in an applied ac field with frequency $\omega=2\pi f$, the frequency dependence of the $\chi'$ and $\chi''$ is significant in a certain temperature range, where $\omega\tau$ is comparable with 1. At any given frequency at low temperature, we have $\omega\tau\gg1$. Increasing of temperature will decrease $\tau$ according to the Arrhenius law. Therefore, $\omega\tau$ decreases to values close to 1 and thus the frequency dependence will be observable. This interpretation is consistent with the more pronounced frequency dependence of the $\chi'$ and $\chi''$ above $T_a$ and close to $T_C$ (Fig. 2).

\subsection{Tuning the pinning strength}

In order to further study the effect of the GBs on the pinning mechanism and confirmation of our interpretation, we tried to tune the pinning strength of the GBs. For this purpose, two samples were prepared at the same preparation conditions; the first one finally sintered in the powder form (henceforth called POW20) and the second one sintered in the bulk (pellet) form (henceforth called BUL20). The sintering of powder produced a coarse powder with typical particle size of 0.8 mm or smaller, which is a result of the attachment of powder particles to each other during the high temperature sintering. For this experiment we used the sample with $x$=0.2 (Pr$_{0.8}$Ag$_{0.2}$MnO$_3$), which has similar physical properties to the sample with $x$=0.15 \cite{17}. The $\chi'$ and $\chi''$ vs T curves, measured in an ac field of 10 Oe, are given in Fig. 4. It can be seen that $T_C$ decreased from 127 K for sample BUL20 to 120 K for sample POW20. This reduction of $T_C$ may be related to the fact that sample BUL20 is more oxygenated than sample POW20, or may be due to the better grain growth in this sample. Interestingly, the anomaly is more pronounced in sample POW20. This behavior is attributed to the weaker pinning strength in sample POW20. Due to the preparation conditions, the intergrain coupling is weaker in sample POW20 than in sample BUL20. Thus because of the improved GBs of sample BUL20, the DWs experience a deeper potential wells which can decrease the depinned walls and affect their movements under the combined effects of thermal activation and ac field. This means that, higher ac field is required to overcome the GBs barriers and move the DWs for a rather large distance in sample BUL20. Fig. 4 clearly shows the role of the GB effects in the occurrence of the anomaly. For sample BUL20, the anomaly is better visible in the $d\chi'/dT$ (inset Fig. 4). Note that, despite the different $T_C$, $T_a$ is almost the same for both samples. The SEM images of the samples (Fig. 5) confirm our interpretation. In sample BUL20, the connectivity between manganite grains is much better than in sample POW20, which results in the enhancement of intergrain coupling.

If just one type of activation energy existed, the internal susceptibility would be expressed by the Debye formula;
\begin{equation}
\chi_{int}=\frac{\chi_0}{1+i\omega\tau}
\end{equation}
where $\chi_0$ is the susceptibility in the static limit ($\omega\tau\ll1$). Equation (1) can be used for a qualitative understanding of the effects of frequency and relaxation time (or temperature) on the susceptibility. Both the internal $\chi'$ and $\chi''$ decrease with increase of $\omega\tau$, and the $\chi''$ has a maximum when $\omega\tau=1$. However in a real case, a very broad distribution of energy barriers should be involved in the pinning of the DWs by the pinning centers like GBs. Kuz'min et al. \cite{20} derived an expression for a rectangular distribution with barriers spanning from a minimum threshold value $E_0$ to a maximum $E_0+W$. The averaged real part of internal susceptibility was obtained as follows;
\begin{equation}
\chi_{int}'=\frac{1}{2}kT\frac{\chi_0}{W}\ln\{1+[\omega\tau_0\exp(\frac{E_0}{kT})]^{-2}\}
\end{equation}
This equation was used to calculate the DW movements contribution to the ac susceptibility \cite{20}. Here, we have used equation (2) to calculate the activation energy of sample BUL20. As it was discussed before, the DW movements dominate the response of the system in low ac fields. So, as shown in Fig. 6, the $\chi'(T)$ data of sample BUL20, measured in a low ac field of 1 Oe, is fitted by equation (2) in the low temperature range. In this temperature range which is below the anomaly temperature, the DWs may jump between the wells. The value of $\tau_0$ is taken equal to 10$^{-9}$ s. In this fit, the demagnetization factor is assumed to be zero. The minimum activation energy $E_0$ was found to be 47 meV for sample BUL20. The $E_0$ value is close to the activation energy ($E_a$) values reported for DyAl$_2$ ($E_a$=40 meV) \cite{14} and Fe$_3$O$_4$ ($E_a$=40 meV) \cite{16}. It is worth to note that the decreasing of the $\chi'$ by increasing the frequency at constant temperature is also supported by equation (2) \cite{20}.

Similar to the sample with x=0.15 (Fig. 1), the $\chi'$ of sample POW20 was measured at different ac fields, before and after that it was well ground manually into a fine powder. As Fig. 7 shows, the ac field dependence of the $\chi'(T)$ is more significant between 90 and 115 K. Due to the weak pinning strength in sample POW20, the anomaly is seen even in an ac field of 7 Oe. From Fig. 7, it can be seen that the ac field dependence is much weaker for the powder form of sample POW20. More importantly, the anomaly almost disappears in the powder form which may confirm that it is not intrinsic in character.

As a result of the above discussion, we may rule out other possible mechanisms such as spin glass as reported for Nd$_{1-x}$Ag$_x$MnO$_3$ \cite{21}. A similar anomaly was observed in La$_{1-x}$MnO$_3$ by Raj Sankar et al. \cite{9}. The authors discussed the nature of the anomaly and attributed it to the DW pinning effects. Nonuniform distribution of Mn$^{4+}$, La and Mn vacancies in the lattice of the self doped compositions, structural distortions, etc., were suggested as the possible pinning centers. However, the results presented here confirm that the anomaly at low temperature in the Ag doped PrMnO$_3$ (Ag + self doped Pr$_{1-y}$MnO$_{3+\delta}$) is originated from the DWs depinning through the GBs. It is important to note that the all GB characteristics (pinning strength, density and type of defects, secondary phase, ...) can affect the motion of the DWs. The other sources of pinning may be active when the DWs move to the interior of grains. However, other factors such as structural transition at low temperatures \cite{9} and residual microstress may also affect the DW movements. Residual microstress is caused by imperfections of various kinds such as dislocations and may hinder the DW motion. Here, it would be useful to know the size of the domains and width of the DWs in manganites. In the ferromagnetic insulator phase (like our samples) of La$_{0.875}$Sr$_{0.125}$MnO$_3$, two types of domains were observed; one is the large plate-shaped domains with micron-size dimensions and the other is the wavy stripe-shaped domains with several hundred nanometers in size \cite{two}. The presence of micron-sized domains has been also reported in other manganites \cite{sci,yu}. In the literature, it has been shown that the DWs are some ten of nanometers thick. The DW width was found to be 38$\pm$10 nm for La$_{0.7}$Ca$_{0.3}$MnO$_3$ thin film \cite{l1}. This is close to the value of 39 nm obtained by Murakami et al. for single crystalline La$_{0.25}$Pr$_{0.375}$Ca$_{0.375}$MnO$_3$ \cite{nature}.

It is interesting to mention that, there is a similarity between the DWs pinning mechanism reported here and flux-line pinning in polycrystalline samples of the high-temperature superconductors (HTSC). There is a broad peak in the $\chi''$ and a drop in the $\chi'$ at a certain temperature below the transition temperature of polycrystalline samples of the HTSC, which are attributed to the GBs. Similar to the DW jumping, the flux creep process occurs at the GBs in low ac fields and become flux flow above an intergrain decoupling field \cite{18}. Shind\'{e} et al. \cite{19} reported that the powdering of the bulk samples of YBa$_2$Cu$_3$O$_{7-\delta}$ and YBa$_2$Cu$_{2.985}$Ag$_{0.015}$O$_{7-\delta}$ leads to the absence of the $\chi''$ peak. It means that the ac losses depend on intergrain coupling and hence occur most likely at the GBs \cite{19}. Field dependence of susceptibility is also suppressed by powdering the sintered samples of the HTSC \cite{ybco}.

\subsection{The effect of thermal cycling}

All of the above discussed measurements were performed during warming run after zero field cooling the samples. In order to gain more insight into the DW dynamics, the $\chi'$ and $\chi''$ of sample POW20 were measured following different thermal cycling (Fig. 8). The $\chi'(T)$ (Fig. 8(a)) and especially $\chi''(T)$ (Fig. 8(b)) data show a considerable amount of thermal hysteresis (TH) below $T_a$ between the field cooled cooling (FCC), field cooled warming (FCW) and zero field cooled warming (ZFCW) runs in an ac field of 10 Oe. Interestingly, the anomaly does not clearly appear during the FCC run in the temperature range of measurement. Similar TH was also observed in the bulk sample with x=0.15 (not shown here). In order to check the effect of temperature rate on TH, each of the runs was performed at two different temperature rates. The higher (lower) temperature rates were approximately 0.48 (0.09), 0.31 (0.04) and 0.37 (0.05) K/min for the FCC, FCW and ZFCW runs, respectively. The observed TH was the same for both selected temperature rates and no temperature rate dependent behavior was found in the runs. In Fig. 8, we have shown the $\chi'$ and $\chi''$ measured at 0.48 (FCC), 0.31 (FCW) and 0.05 K/min (ZFCW).

In general, TH in ac susceptibility measurements has been shown  to be a characteristic feature of a first order phase transition (ferromagnetic - antiferromagnetic) \cite{22}. This phenomenon is observed in phase-separated manganites and is more pronounced at around $T_N$ \cite{23}. Such TH is not destroyed by powdering the bulk sample as reported for phase-separated Pr$_{0.5}$Sr$_{0.5}$MnO$_3$ \cite{23}. However, as it is shown in Fig. 7, the anomaly in our samples is an extrinsic effect and can not be related to a first order phase transition. Hence, the above experiments bring out the role of the DWs for explaining the observed TH in the $\chi'$ and $\chi''$. Appearance of TH below $T_a$, indicates thermally irreversible DW pinning and depinning. In fact, the metastable positions of the DWs are strongly dependent on the temperature, applied field and thermal cycling. As a consequence, domain size and/or DWs configuration are not the same for different thermal cycling. In the ZFCW run, when temperature increases to $T_a$, combined effects of ac field and temperature lead to the depenning of the DWs. This process is not a thermally reversible process since during long range wall movements the potential barriers of the GBs must be overcome. In the FCC run, the magnetic domains are formed as soon as the sample is cooled below $T_C$. With decreasing temperature, the $\chi'$ and $\chi''$ tend to decrease due to the decrease of the DWs response to the applied ac field. When temperature decreases, the DWs are unable to fully reverse their motion back to their original positions at the same temperature in the ZFCW run. In fact, temperature plays a role similar to a dc magnetic field. In the FCW run, the $\chi'$ and $\chi''$ magnitudes are larger and TH area is smaller (relative to the FCC run) than the ZFCW run. This means that in the FCW run, the DWs are more active, the effective potential barrier is lower and the ac loss is larger than that of ZFCW run. It is important to note that the ac loss is present in an ac field of 1 Oe because of the DWs jumping from well to well. However, the TH does not occur in such low ac fields (Fig. 8). It means that the TH is correlated with the long range DW movements through the GBs.

We now inspect the frequency dependence in the FCC and FCW runs. Figs. 9 and 10 show such measurements for sample POW20 in an ac field of 10 Oe. As can be seen in both figures, the $\chi'$ and $\chi''$ are divided into two parts. At temperatures below 98 K, the $\chi'$ and $\chi''$ are practically frequency independent. However, in the temperature range between 98 and 115 K, the data exhibit frequency dependent behavior. It means that, there is no frequency dependence in the region of TH. As it was discussed before, frequency dependence of the $\chi'$ and $\chi''$ occurs when $\omega\tau\sim1$. At low temperatures, the average relaxation time is very long and it decreases drastically with increasing temperature. In the temperature range between 98 and 115 K, $\omega\tau$ decreases to values not far from 1 and decreasing of the $\chi'$ and $\chi''$ with frequency is clearly observable. In this temperature range, the appearance of the $\chi''$ peak is accompanied with a rapid rise in the $\chi'$. A higher frequency will delay the decrease of $\omega\tau$ with respect to temperature, so that the $\chi''$ peak shifts to the higher temperatures.

All the above discussions indicate that the low temperature magnetic state of the Ag doped PrMnO$_3$ is inhomogeneous and is not in a state of global minimum. The results of this study provide a qualitative understanding of the roles of DWs and GBs in the inhomogeneous state of manganites. However, in order to better understand the complicated magnetic behaviors of the Ag doped PrMnO$_3$, more detailed studies are still needed.

\section{Conclusions}

The real and imaginary parts of ac susceptibility of Pr$_{1-x}$Ag$_x$MnO$_3$ ($x$=0.15, 0.20) exhibit unusual and interesting behaviors. Ac field, frequency, temperature, thermal cycling and grain coupling dependencies of the $\chi'$ and $\chi''$ were discussed within the framework of domain wall dynamics. The results indicate that the grain-boundaries have a crucial effect on the domain wall dynamics in this compound. The main results are as follows: (i) An anomaly was observed above a threshold ac field at low temperatures. The anomaly was found to be dependent on the intergrain coupling and was attributed to the cooperative depinning of the domain walls from the grain-boundaries accompanying with a rather large distance movements. (ii) Powdering the sintered bulk or sintered powders leads to the disappearance of the anomaly which may indicate that the anomaly is not intrinsic. (iii) Significant thermal hysteresis was observed in the $\chi'$ and $\chi''$ below the anomaly temperature in an ac field of 10 Oe. We argued that this thermal hysteresis indicates thermally irreversible domain wall pinning and depinning which brings out the nonequilibrium configuration of the domain walls. (iv) The $\chi'$ and $\chi''$ were practically frequency dependent (independent) at temperatures above (below) the anomaly temperature. The effect of frequency was discussed by considering the temperature dependence of the relaxation process of the domain walls.

\section*{ACKNOWLEDGMENTS}

This work was supported by Isfahan University of Technology (IUT). The authors would like to thank Dr. S. J. Hashemifar for his help in preparing the manuscript.

\newpage
\begin{figure}[t]
\includegraphics*{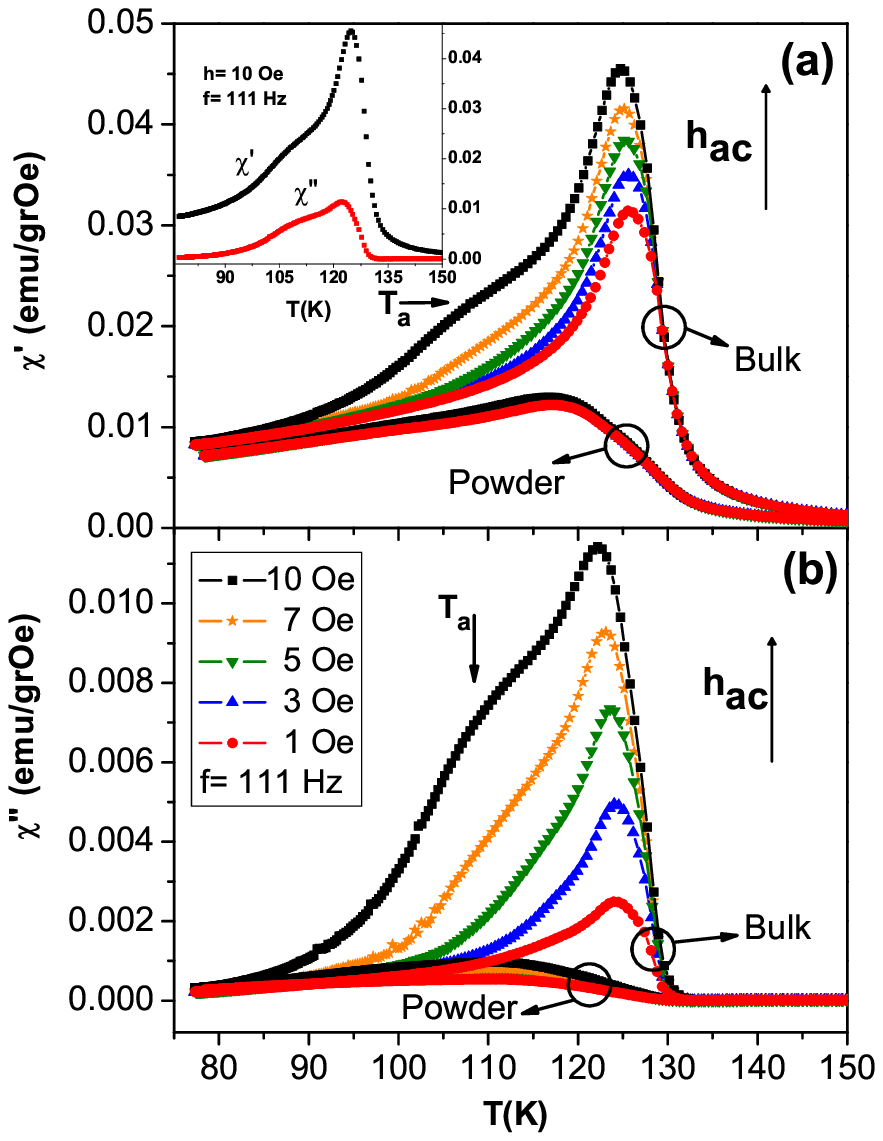}
\caption{The real (a) and imaginary (b) parts of ac susceptibility at different ac fields (1, 3, 5, 7, 10 Oe) for the bulk and powder samples of Pr$_{0.85}$Ag$_{0.15}$MnO$_3$. Inset shows the real and imaginary parts for the bulk sample in an ac field of 10 Oe.}
\end{figure}
\begin{figure}[t]
\includegraphics*{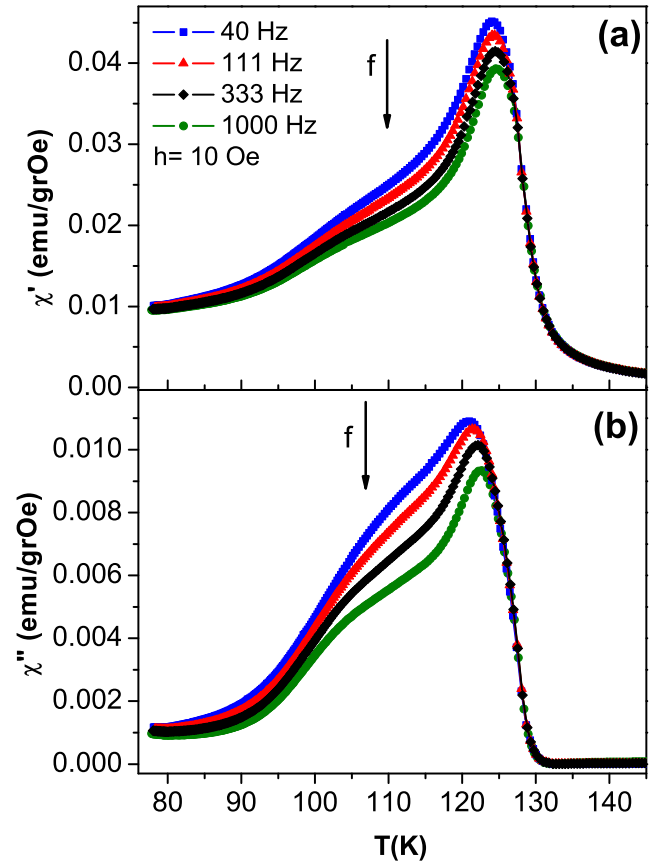}
\caption{Temperature dependence of the real (a) and imaginary (b) parts of ac susceptibility at different frequencies (40, 111, 333, 1000 Hz) in an ac field of 10 Oe for the bulk sample of Pr$_{0.85}$Ag$_{0.15}$MnO$_3$.}
\end{figure}
\begin{figure}[t]
\includegraphics[width=9cm]{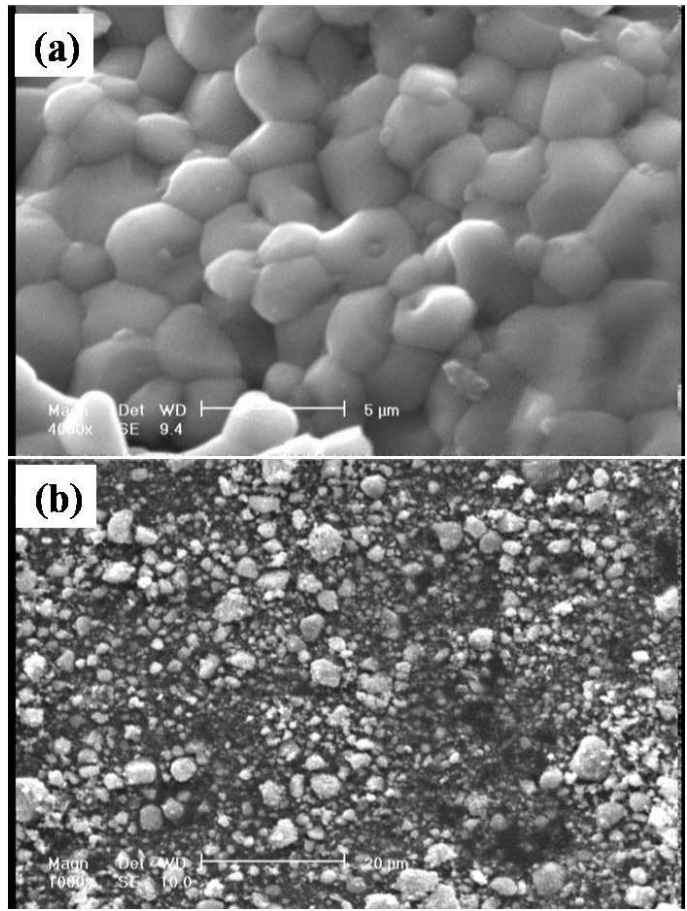}
\caption{The SEM images of the Pr$_{0.85}$Ag$_{0.15}$MnO$_3$, (a) bulk sample (b) powder sample. The scale bars are 5 and 20 $\mu m$ respectively.}
\end{figure}
\begin{figure}[t]
\includegraphics*{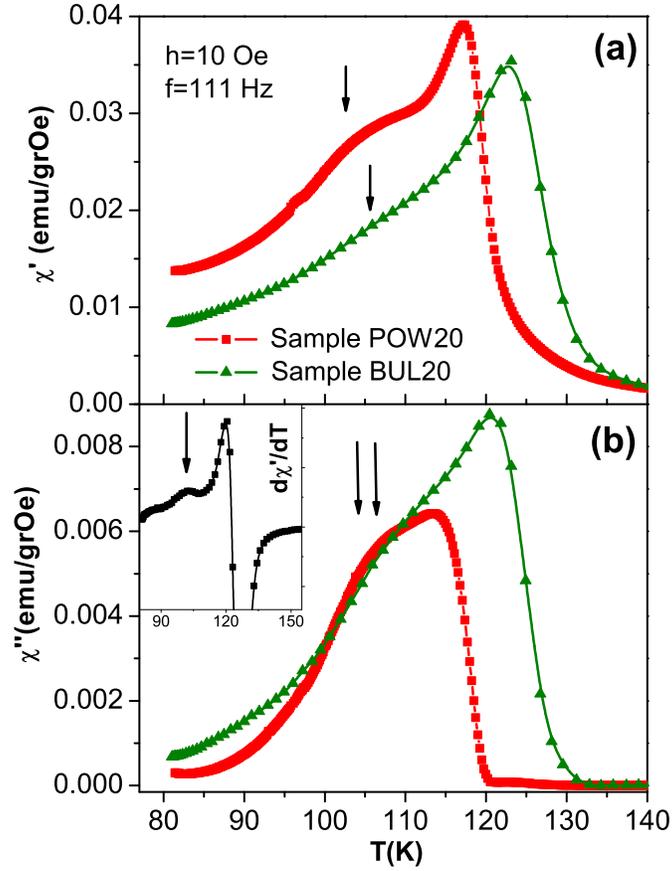}
\caption{Temperature dependence of the real (a) and imaginary (b) parts of ac susceptibility in an ac field of 10 Oe for samples BUL20 and POW20. Inset shows the d$\chi'$/dT for sample BUL20. The arrows indicate the anomaly.}
\end{figure}
\begin{figure}[t]
\includegraphics[width=9cm]{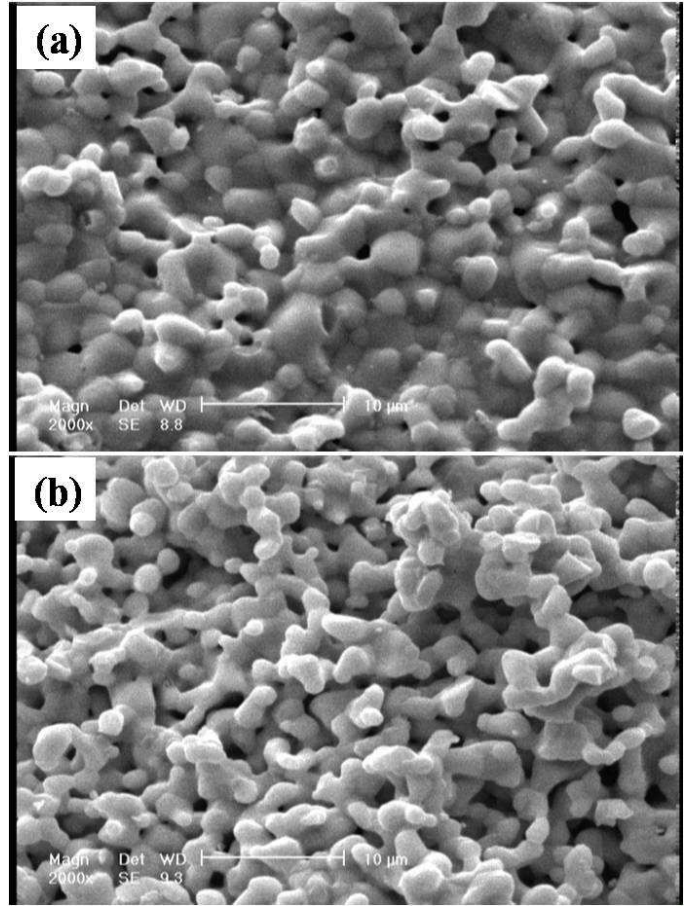}
\caption{The SEM images of sample BUL20 (a) and sample POW20 (b). The scale bars are 10 $\mu m$.}
\end{figure}
\begin{figure}[t]
\includegraphics*{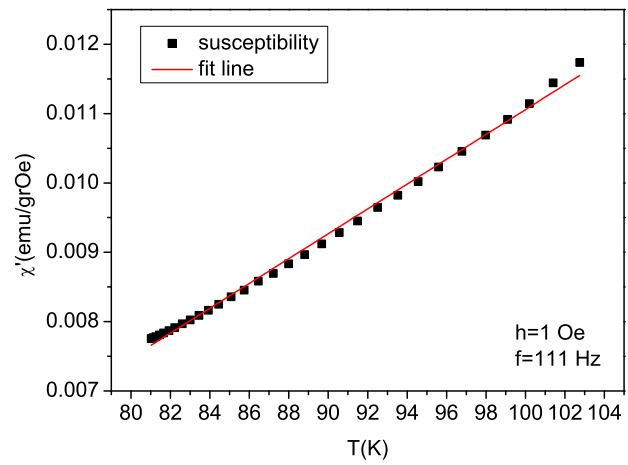}
\caption{Fit of the low temperature susceptibility data of sample BUL20 by equation 2.}
\end{figure}
\begin{figure}[t]
\includegraphics*{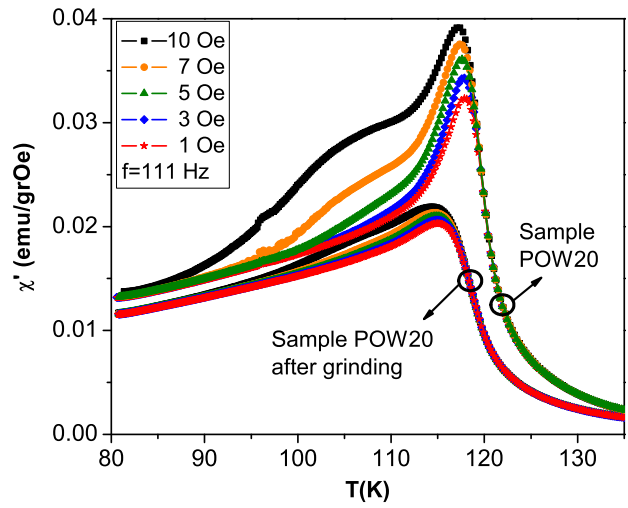}
\caption{The real part of ac susceptibility at different ac fields (1, 3, 5, 7, 10 Oe) for sample POW20 before and after grinding.}
\end{figure}
\begin{figure}[t]
\includegraphics*{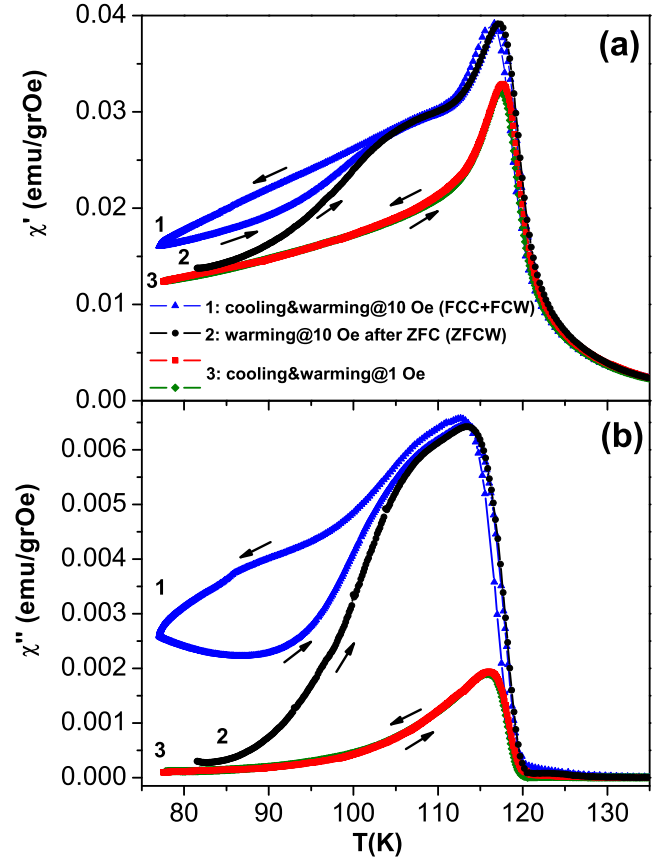}
\caption{The real (a) and imaginary (b) parts of ac susceptibility for sample POW20. The data were measured during field cooled cooling (FCC), field cooled warming (FCW) and zero field cooled warming (ZFCW) runs.}
\end{figure}
\begin{figure}[t]
\includegraphics*{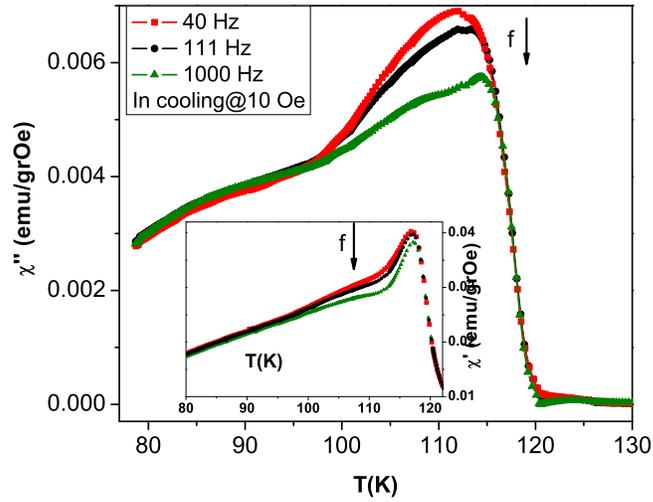}
\caption{Temperature dependence of the imaginary part of ac susceptibility at different frequencies (40, 111, 1000 Hz) for sample POW20. The data were measured during cooling the sample in an ac field of 10 Oe. Inset shows the corresponding data of the real part.}
\end{figure}
\begin{figure}[t]
\includegraphics*{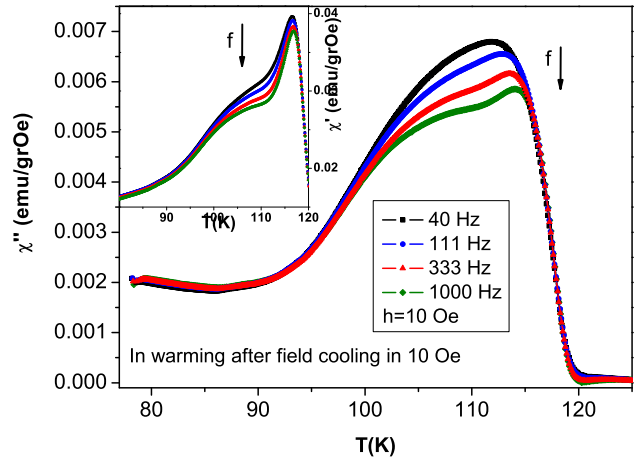}
\caption{Temperature dependence of the imaginary part of ac susceptibility for sample POW20 measured at different frequencies (40, 111, 333, 1000 Hz) in an ac field of 10 Oe. The data were measured during warming after field cooling the sample in 10 Oe. Inset shows the corresponding data of the real part.}
\end{figure}
\end{document}